\documentclass[twocolumn,showpacs,amsmath,amssymb,prl]{revtex4}

\usepackage{graphicx}
\usepackage{bm}

\begin{document}

\title{Loss of
Andreev Backscattering in Superconducting Quantum Point Contacts}
\author{N. B. Kopnin $^{(1,2)}$}
\author{A. S. Mel'nikov $^{(3,4)}$}
\author{V. M. Vinokur $^{(4)}$}
\affiliation{$^{(1)}$ Low Temperature Laboratory, Helsinki
University of
Technology, P.O. Box 2200, FIN-02015 HUT, Finland,\\
$^{(2)}$ L. D. Landau Institute for Theoretical Physics, 117940
Moscow, Russia\\
 $^{(3)}$ Institute for Physics of Microstructures,
Russian Academy of Sciences, 603950, Nizhny Novgorod, GSP-105,
 Russia,\\
$^{(4)}$ Argonne National Laboratory, Argonne, Illinois 60439 }

\date{\today}

\begin{abstract}
We study effects of magnetic field on the energy spectrum in a
superconducting quantum point contact. The supercurrent induced by
the magnetic field leads to intermode transitions between the
electron waves that pass and do not pass through the constriction.
The latter experience normal reflections which couple the states
with opposite momenta inside the quantum channel and create a
minigap in the energy spectrum that depends on the magnetic field.
\end{abstract}

\maketitle

The transport in superconducting--normal-metal (SN) mesoscopic
hybrid structures is controlled by normal and Andreev reflections.
The fundamental property of the Andreev reflection is its almost
exact backscattering: trajectories of an incoming particle and
reflected hole coincide within an angle of a maximum $\sim (k_F\xi
)^{-1}$ where $\xi $ is the superconducting coherence length and
$k_F$ is the Fermi wave vector. Backscattering holds also in the
presence of magnetic field \cite{galaiko} since the electron-hole
conversion length $\sim\xi$ is much smaller than the Larmor radius
$r_L$, i.e., $\xi/r_L\sim (B/H_{c2}) (k_F\xi)^{-1}$. This has been
checked, for example, by the electron focusing technique
\cite{tsoi}. Deviation from backscattering come from interference
of electron and hole waves with an inhomogeneous order parameter
phase. Effects of a supercurrent-induced transverse force on
backscattering have been considered in Refs.\
\cite{kummel-v,Cook}. In Refs.\ \cite{interferometer} the phase
difference introduced between the partial waves resulted in a
decrease in the reflected hole intensity leading to an interplay
between the normal and Andreev processes.

Though deviation from exact backscattering is small, it can still
be noticeable if it is comparable with the size of the setup. A
convenient device that satisfies this requirement is a ballistic
superconducting quantum point contact. In the present paper we
study how the violation of the fundamental semiclassical
backscattering property of the Andreev reflection affects the
subgap energy spectrum in such contact. Under the condition of
exact backscattering, each electron that passes through the
constriction is reflected as a hole that returns along the same
trajectory and thus has also to pass through the constriction. The
energy spectrum of subgap states is then $E=\pm\Delta_0\cos\chi$
where $\Delta_0$ is the superconducting gap and $2\chi$ is the
phase difference between the electrodes \cite{kulik-om1,been1}.
The two energy branches which correspond to quasiparticles
propagating through the constriction in opposite directions cross
the Fermi level $E=0$ at $2\chi =\pi$.   We show that the loss of
exact backscattering leads to a dramatic change in the spectrum
such that a minigap appears near $2\chi =\pi$. The deviation from
backscattering can be produced, for example, by an exchange,
during the Andreev process, of a Cooper pair momentum induced by
an applied magnetic field. This momentum mixes the channel modes
with the modes that do not penetrate inside but are normally
reflected from the channel end. The normal reflections couple the
waves propagating through the constriction in the opposite
directions and lead to formation of a minigap in the energy
spectrum similar to that for contacts with normal scatterers
\cite{bagwellimp,haber,zaitsev,been2}. Varying the magnetic field
one can tune the degree of normal reflection and manipulate the
minigap thus controlling the transport properties of the contact.

\begin{figure}[t]
\centerline{\includegraphics[width=1.0\linewidth]{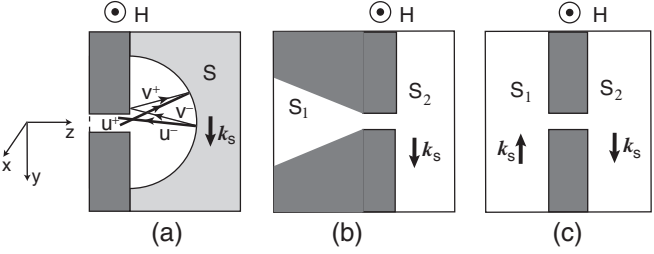}} %
\caption{(a) A single-mode channel is open to a normal region
(white semicircle) in a contact with a superconductor (grey
region). Andreev reflected trajectories deviate from initial
direction due to the transverse pair momentum $\hbar {\bf k}_s$
and experience normal reflections from the insulator surface
(black), which couple right-moving $u^+$ and left-moving $u^-$
states. (b) Asymmetric and (c) symmetric point contacts.}
\label{fig1}
\end{figure}

{\it Model.}-- The loss of exact backscattering at the Andreev
reflections in a quantum point contact can be more clearly
illustrated for an auxiliary structure shown in Fig.\
\ref{fig1}~(a): A single mode channel with a radius $a\sim
k_F^{-1}$ is open into a normal semi-spherical region with a
radius $R$ much larger than the superconducting coherence length
$\xi$. The normal region is surrounded by a superconductor which
carries a supercurrent with a momentum $\hbar {\bf k}_s$
perpendicular to the channel axis. For $R\gg \xi$ one can describe
quasiparticle propagation using a trajectory representation. Due
to the transfer of $\hbar {\bf k}_s$ the Andreev reflected
trajectory deviates from its initial direction \cite{kummel-v}
such that it can miss the constriction and experience normal
reflections from the insulating barrier. The trajectory returns to
the constriction after several reflections, which results in
coupling of states propagating through the constriction in the
opposite directions. The transfer of ${\bf k}_s$ causes a
trajectory deflection by an angle $k_s/k_F$, thus the shift of the
trajectory over a distance $R$ would be $k_sR /k_F$. The
probability of normal reflection thus depends on the ratio of the
trajectory shift to the transverse channel dimension $a$. For a
single-mode quantum channel $a\sim k_F^{-1}$, this ratio is
$k_sR/k_Fa\sim k_sR$.

In a superconducting point contact, the quasiparticle wave
functions for subgap states decay over distances of the order of
$R\sim \xi$, which gives an estimate $k_sR\sim k_s\xi \lesssim 1$.
However, the trajectory shift $k_s\xi /k_F$ is less than the wave
length so that the trajectory picture is not applicable. Moreover,
a single-mode contact with a radius $a\sim k_F^{-1}$ radiates an
electronic wave which is determined by diffraction. To match the
particle $u^\pm_0$ and hole $v^\pm _0$ waves propagating along the
channel with the quasiparticle waves in the corresponding
superconductors we assume, for simplicity, that the right
superconductor occupies a halfspace $z>0$ and has a specularly
reflecting surface. We introduce the spherical coordinates
$x=r\sin\theta\cos\phi, y=r\sin\theta\sin\phi, z=r\cos\theta$ and
expand the wave functions in spherical functions with odd angular
momenta. This ensures vanishing of the microscopic wave function
at the superconducting/insulator boundary $\theta =\pi /2$. Far
from the origin ($r\gg a$) we have
\begin{equation}
\left(\begin{array}{c}u^\pm \\ v^\pm \end{array} \right)=
\frac{e^{\pm ik_Fr}}{r} \sum _{l=2n+1,m} P_{l,m}(\theta,
\phi)\left(\begin{array}{c}U^\pm _{l,m}(r)\\ V^\pm
_{l,m}(r)\end{array}\right) \ . \label{spherical}
\end{equation}
Different angular harmonics are orthogonal to each other within
$0<\theta <\pi/2$. For a waveguide $a\lesssim k_F^{-1}$, the
radiated/incident diffraction field is $\exp(\pm
ik_Fr)\cos\theta/r$. We now  assume that it is this only mode
proportional to $P_{1,0}=\cos\theta$ that ideally transforms into
the channel mode $u_0$, $v_0$ without reflections, while all other
modes $l\ne 1$ are normally reflected from the waveguide end
without transmission into the channel.

{\it Scattering Matrix.}-- For more quantitative description we
use the Bogolubov - de Gennes (BdG) equations:
\begin{eqnarray}
\left[\frac{1}{2m}\left(-i\hbar {\bm \nabla} -\frac{e}{c}{\bf
A}\right)^2
-E_F\right] u+\Delta v&=&E u \ , \label{Beq1}\\
\left[\frac{1}{2m}\left(-i\hbar {\bm \nabla} +\frac{e}{ c} {\bf
A}\right)^2 -E_F\right] v-\Delta ^* u&=&-E v \ , \label{Beq2}
\end{eqnarray}
where ${\bf A}$ is the vector potential of the magnetic field
${\bf B}=B(z)\hat{\bf x}$, and $\Delta$ is the gap function. We
assume a step-like form $\Delta =\Delta_0 e^{i(\chi + {\bf k}_s
{\bf r})}$, where $\chi _{R,L}=\pm \chi $ is the zero-field
constant phase in the right (left) superconductor with $2\chi$
phase difference between the superconductors; ${\bf k}_s$ is a
constant wave vector. It enters the superconducting velocity $
m{\bf v}_s=\hbar {\bf k}_s-(2e/c){\bf A} $ that determines the
difference in the eikonals of particle and hole wave functions,
$(m/\hbar)\int {\bf v}_s \cdot d {\bf r}$. The magnetic field is
screened in the superconductor, ${\bf A}=\lambda_L B_0 \hat{\bf
y}(\exp(-z/\lambda_L)-1)$, at the London length $\lambda_L$. From
the screening condition ${\bf v}_s=0$ at large $z$ we find ${\bf
k}_{sR}=(2\pi/\phi_0)\lambda_L B_0 \hat{\bf y}$, where $\phi_0$ is
the flux quantum. Assuming for simplicity $\lambda _L\gg \xi $ we
neglect ${\bf A}$ in the region $r\sim \xi$ where the wave
functions are localized. The parameter $k_s\xi $ that determines
the probability of normal reflections of the channel modes is
$k_s\xi \sim B_0/H_{cm}$ where $H_{cm}\sim \phi_0/(\lambda_L\xi)$
is the thermodynamic critical field of the superconductor. The gap
$\Delta_0$ is suppressed near the superconductor surface. However,
this does not change the backscattering properties of Andreev
reflection; we ignore it in what follows.

Inside the single-mode channel there are two particle and two hole
waves $\propto e^{\pm ik_z z}$ with amplitudes $u_0^\pm$ and
$v_0^\pm$, respectively, corresponding to the momentum projections
$\pm \hbar k_z$ on the $z$ axis. A particle $u^+_0$ and a hole
$v^-_0$ propagate in the $+z$ direction while a particle $u^-_0$
and a hole $v^+_0$ propagate in $-z$ direction. Using the scheme
employed in Ref.\ \cite{been2}, we introduce the scattering
matrices $\hat S_R (\epsilon,\chi,{\bf k}_{sR})$ and $\hat
S_L(\epsilon,-\chi,{\bf k}_{sL})$ that relate the incident and
outgoing wave amplitudes respectively at the right, $z=d/2$, and
left, $z=-d/2$ ends of the channel:
\begin{equation}
\left(\begin{array}{c}u_0^-\\
v_0^+\end{array}\right)_{R}\! =\hat
S_R\left(\begin{array}{c}u_0^+\\
v_0^-\end{array}\right)_{R}  , \;
\left(\begin{array}{c}u_0^+\\
v_0^-\end{array}\right)_{L}\! = \hat S_L
\left(\begin{array}{c}u_0^-\\
v_0^+\end{array}\right)_{L}  . \label{right/left}
\end{equation}
Here $d\ll\xi$ is the channel length, ${\bf k}_{sL}$ and ${\bf
k}_{sR}$ are the superflow momenta in the left and right
superconductors, respectively. The wave functions at the both ends
of the channel have different phase factors:
\begin{equation}
\left(\begin{array}{c}u_0^{\pm}\\ v_0^\pm
\end{array}\right)_{R}=e^{\pm ik_z d} \left(\begin{array}{c}
u_0^{\pm}\\ v_0^\pm \end{array}\right)_{L} \ . \label{phases}
\end{equation}
The solvability condition of Eqs.\ (\ref{right/left}) and
(\ref{phases}) yields
\begin{equation}
{\rm det}\left(1-e^{i\hat \sigma_z k_z d}\hat S_Re^{i\hat
\sigma_zk_z d}\hat S_L\right)=0 \ . \label{zero-det}
\end{equation}

The matrix $\hat S$ is unitary: $\hat S\hat S^\dagger =1$. Indeed,
the BdG equations (\ref{Beq1}), (\ref{Beq2}) conserve the
quasiparticle flow
\begin{eqnarray*}
{\rm div}\left[u^*\left(-i\hbar {\bm \nabla} -\frac{e}{c}{\bf
A}\right)u +u \left(i\hbar {\bm \nabla} -\frac{e}{c}{\bf
A}\right)u^*\right. \\
\left. -v^*\left(-i\hbar {\bm \nabla} +\frac{e}{c}{\bf A}\right)v
-v\left(i\hbar {\bm \nabla} +\frac{e}{c}{\bf A}\right)v^*\right]
=0 \ .
\end{eqnarray*}
Since this flow vanishes for $|E| <\Delta_0$ deep in the
superconductor it should be zero also in the channel, whence
$|u^+_0|^2+|v^-_0|^2=|u^-_0|^2+|v^+_0|^2$ which results in $\hat
S\hat S^\dagger =1$.

We now calculate the matrix $S$ explicitly with the account of
reflections from the channel end. Wave functions decaying into the
right superconductor obey the relations
\begin{eqnarray}
v_R^+ &=&e^{-\frac{i}{2}(\chi +{\bf k}_{sR} {\bf r})}\check a
^{+}_\epsilon ({\bf k}_{sR}) e^{-\frac{i}{2}(\chi +{\bf k}_{sR}
{\bf r})}u_R^+ \ , \label{andr1}
\\
u_R^- &=&e^{\frac{i}{2}(\chi +{\bf k}_{sR} {\bf r})}\check a
^{+}_\epsilon ({\bf k}_{sR}) e^{\frac{i}{2}(\chi +{\bf k}_{sR}
{\bf r})}v_R^- \ , \label{andr2}
\end{eqnarray}
that couple the electron and hole amplitudes near the channel end
$|{\bf r}|\ll\xi$. Here
\[
\check a ^{\pm}_\epsilon ({\bf k}_s)=\epsilon+i\xi{\bf k}_{s} {\bm
\nabla}/2k_F \mp i\sqrt{1-(\epsilon+i\xi{\bf k}_{s} {\bm
\nabla}/2k_F)^2} \ ,
\]
$\xi =\hbar v_F/\Delta_0$, and $\epsilon=E/\Delta_0$. For the left
superconductor, similar expressions hold  with $\chi \rightarrow
-\chi$ and $\check a^+\rightarrow \check a^-$.


Let us consider the matrix $\hat S$ at the right end of the
channel. We place the origin of the coordinate system at the right
channel end and consider the case $\chi =0$ since, generally, $
\hat S(\chi)= e^{i\chi\sigma_z/2}\hat
S(\chi=0)e^{-i\chi\sigma_z/2} $. According to Eq.\
(\ref{spherical}), the wave function amplitudes in the right
superconductor at distances $a \ll r \ll\xi$ from the origin have
the form (we omit the index $R$)
\begin{eqnarray*}
U^+ = u_0^+ P_{1,0} + \Psi _u^+  &,&
V^+ = v_0^+ P_{1,0} + \Psi _v ^+  \ ,\\
U^- = u_0^- P_{1,0} + \Psi_u &,&  V^- = v_0^- P_{1,0} + \Psi_v  \
.
\end{eqnarray*}
The amplitudes $\Psi _u$ and $\Psi _v$ stand for the sum of
components with $l\ne 1$ in Eq.\ (\ref{spherical}) and describe
the modes which experience normal reflections at the channel end.
Introducing the operator $\check R_\epsilon$ that describes this
reflection we can write the relation between the amplitudes $ \Psi
_u^+ =\check R_\epsilon \Psi_u $, $ \Psi _v ^+ =\check
R_{-\epsilon} \Psi_v $. The functions $\Psi_u, \Psi_v$ are
orthogonal to the mode $P_{1,0}$: $\langle P_{1,0} |
\Psi_{u,v}\rangle=0$. The angular brackets denote the angular
average within $0<\theta <\pi /2$.

The modes $\Psi _u$ and $\Psi _v$ together with the traversing
mode $u_0$, $v_0$ experience Andreev reflections while only $u_0$
and $v_0$ contribute to the flow through the channel. The
unitarity $\hat S \hat S^\dagger =1$ implies that the
quasiparticles which are scattered normally off the superconductor
surface and the channel end will eventually return into the
constriction either as particles or as holes after certain number
of Andreev reflections at the superconducting side. The Andreev
relations (\ref{andr1},\ref{andr2}) should be applied to the total
functions $U$ and $V$ at a hemisphere with the center at the
origin and the radius much smaller than $\xi$ where ${\bf k}_s
{\bf r}\sim k_s r\sin\theta\sin\phi\ll 1$. Taking the derivatives
only of the rapidly varying exponents in Eq.\ (\ref{spherical}) we
obtain
\begin{eqnarray}
v_0^+P_{1,0}+\check R_{-\epsilon}\Psi_v
&=&e^{-i\varphi_\epsilon}(u_0^+P_{1,0}+\check R_{\epsilon}\Psi_u) \ ,
\label{aux1}\\
u_0^-P_{1,0}+\Psi_u
&=&-e^{i\varphi_{-\epsilon}}(v_0^-P_{1,0}+\Psi_v) \ , \label{aux2}
\end{eqnarray}
where
\begin{equation}
e^{i\varphi_\epsilon}= \epsilon-\xi k_{sr} /2
 +i\sqrt{1-(\epsilon- \xi k_{sr}/2)^2} \ ,
 \label{exp}
\end{equation}
and $k_{sr}= k_s\sin\theta\sin\phi$. Since the normal reflection
at the channel end is associated with the momentum transfer of the
order of $\hbar k_F$ one can neglect the dependence of $\check R$
on energy on the scale $\Delta _0$ and assume $\check R
=-e^{i\varphi_r}$ where $\varphi_r$ is a constant phase shift. We
then solve Eqs.\ (\ref{aux1}), (\ref{aux2}) for the functions
$\Psi_u, \Psi_v$ and use the orthogonality $\langle P_{1,0} |
\Psi_{u,v}\rangle=0$. This yields two equations that couple the
amplitudes $u_0^+,v_0^-$ to $ u_0^-,v_0^+$ through the matrix
\begin{equation}
\hat S =\frac{1}{1-c_\epsilon^2}
\left(e^{-i\varphi_r\hat\sigma_z}(|c_\epsilon|^2-1)-
(c_{\epsilon}-c^*_{\epsilon})e^{i\chi\hat \sigma_z}\hat \sigma_x
 \right)  \label{Smatrix}
\end{equation}
where
\begin{equation}
c_\epsilon=\frac{\langle
P_{1,0}\left(1+e^{i(\varphi_{-\epsilon}-\varphi_{\epsilon})}\right)^{-1}
P_{1,0}\rangle} {\langle
P_{1,0}\left(e^{i\varphi_{\epsilon}}+e^{i\varphi_{-\epsilon}}\right)^{-1}
P_{1,0}\rangle} \ .
\end{equation}
One sees from Eq.\ (\ref{exp}) that
$e^{i\varphi_\epsilon}(k_s)=-e^{-i\varphi_{-\epsilon}}(-k_s)$
whence $c^*_\epsilon(k_s)=-c_{-\epsilon}(-k_s)$. Moreover,
$c_\epsilon(k_s)$ is an even function of $k_s$ because a change in
the sign of $k_s$ can be compensated by the shift $\phi\rightarrow
\pi +\phi$ in the integral over the angles. Thus,
$c^*_\epsilon(k_s)=-c_{-\epsilon}(k_s)$. For small $k_s\xi$,
\begin{equation}
c_\epsilon \simeq  e^{i\eta}-\frac{i(k_s\xi)^2\langle
P_{1,0}^2\sin^2\theta\sin^2\phi\rangle e^{2i\eta}}{8\langle
P_{1,0}^2\rangle\sin^3\eta} \ . \label{small-H}
\end{equation}
Without $k_s$ one has $c_\epsilon =e^{i\eta}$ where $
e^{i\eta}=\epsilon+i\sqrt{1-\epsilon^2}$. As a result, the
diagonal components of $\hat S$ vanish, thus the $+p_z$ and $-p_z$
states are decoupled.

In the diffraction picture, the transitions that couple the
penetrating and non-penetrating modes are caused by the
angle-dependent Doppler shift of energy proportional to $k_{sr}$
in Eq.\ (\ref{exp}). As a result, the wave fronts of reflected
holes are distorted as compared to those of incident particles.
The interference of these waves near the channel end results in
the suppression of the amplitude of the Andreev reflected wave
entering the channel.

{\it Results.}-- Consider first the zero-bias conductance of a
normal-metal -- quantum-channel--superconductor junction
\cite{BTK} $ G_s= (e^2/\pi\hbar ) (1-|S_{11}|^2+|S_{12}|^2) $
where $|S_{11}|^2$ and $|S_{12}|^2$ are probabilities of normal
and Andreev reflection, respectively. We get for small $k_s\xi$
\[
 G_s= \frac{ e^2}{\pi\hbar}
\left[2-\frac{2(|c_\epsilon|^2-1)^2}{|c_\epsilon^2-1|^2}\right]
_{\epsilon=0} \simeq \frac{ e^2}{\pi\hbar}
 \left[2-\frac{1}{2}\left(\frac{B_0}{H_c}\right)^4\right] \ .
\]
Here we introduce a field $H_c\sim H_{cm}$ through
\begin{equation}
\frac{B_0^2}{H_c^2}=\frac{(k_s\xi)^2 \langle
P_{1,0}^2\sin^2\theta\sin^2\phi\rangle}{4\langle
P_{1,0}^2\rangle}=\frac{(k_s\xi)^2}{20} \ . \label{Hc}
\end{equation}

Consider now an asymmetric structure that consists of a
superconducting tip with a curvature radius smaller than $\lambda
_L$ in a contact with a bulk superconductor, see Fig.\
\ref{fig1}~(b). In this case ${\bf k}_{sL}=0$ while ${\bf
k}_{sR}={\bf k}_{s}\ne 0$. On the right end of the channel the
matrix $\hat S_R=\hat S(\epsilon,\chi,{\bf k}_s)$ is determined by
Eq.\ (\ref{Smatrix}). On the left end the matrix $\hat S_L=\hat
S(\epsilon,-\chi,0)$ assumes an Andreev form $ \hat S_L=e^{-i\eta}
e^{-i\chi\sigma_z}\hat \sigma_x$. The phase shift $k_zd-\varphi
_r$ drops out and Eq.\ (\ref{zero-det}) yields
\begin{equation}
\left(1-c_\epsilon ^2\right)e^{i\eta}-\left(1-c_\epsilon
^{*2}\right)e^{-i\eta}=2\left(c_\epsilon ^*-c_\epsilon \right)
\cos (2\chi ) \ . \label{eq-asym}
\end{equation}

For $k_s=0$ and $c_\epsilon =e^{i\eta}$ we obtain a standard
gapless expression $\epsilon=\pm\cos\chi$. For a nonzero $k_s$, a
gap opens in the energy spectrum. Indeed, consider Eq.\
(\ref{eq-asym}) in the limit of small magnetic fields and
energies. It becomes
\[
\epsilon ^2 =\cos ^2\chi +\frac{1}{8}\left[\left(ic_{\epsilon}
+1\right)^2+\left(ic^*_{\epsilon} -1\right)^2\right]_{\epsilon =0}
\ .
\]
Within the leading terms in $B/H_c$ we find
\begin{equation}
\epsilon^2 = \cos^2\chi + \epsilon _g^2 \ . \label{spectrum1}
\end{equation}
where the minigap in the spectrum is
$\epsilon_g=\frac{1}{4}(B_0/H_{c})^2$.

In the case of a symmetric contact shown in Fig.\ \ref{fig1}~(c)
the solution of the screening problem yields ${\bf k}_{sL}=-{\bf
k}_{sR}=-{\bf k}_{s}$. The spectral equation (\ref{zero-det}) with
$\hat S_R=\hat S(\epsilon,\chi,{\bf k}_s)$ and $\hat S_L=\hat
S(\epsilon,-\chi,-{\bf k}_s)$ reduces to
\begin{equation}
\frac{(c_\epsilon +c_{\epsilon}^*)^{2}}{4}=\cos^2\chi +\frac{
(|c_\epsilon|^2-1)^2}{(ic_\epsilon -ic_{\epsilon}^*)^{2}}
\sin^2(k_zd-\varphi_r)\ . \label{eq-symm}
\end{equation}
The spectrum has a gap in the presence of $k_s$. In the limit of
low energies and small magnetic fields the right hand side of Eq.\
(\ref{eq-symm}) can be treated as a perturbation. We put
$(|c_\epsilon|^2-1)^2\simeq(B_0/H_{c})^4$ where $H_c$ is defined
in Eq.\ (\ref{Hc}) while $(c_\epsilon -c_{\epsilon}^*)^2\approx
-4$. At the same time, $(c_\epsilon +c_{\epsilon}^*)^2/4\simeq
\epsilon^2$. Finally we get Eq.\ (\ref{spectrum1}) where
\begin{equation}
\epsilon_g=\frac{1}{2}(B_0/H_{c})^2\left|\sin(k_zd-\varphi_s)\right|
\ . \label{gap-sym}
\end{equation}

{\it Discussion.}-- Since the wave vector $k_s\lesssim \xi^{-1}$
is much smaller than $k_F$ it induces transitions only between the
modes with close transverse quantum numbers. Thus, the predicted
effect can be more easily seen for a contact that is transparent
only for a few modes. On the contrary, for a multi-mode channel,
the coupling to the reflected modes that mixes ${\bf p}$ and
$-{\bf p}$ states has a small weight while the transitions occur
mostly between the penetrating modes. For a large area SNS
junction these transitions result in the subgap spectrum
instability with formation of energy bands \cite{galaiko}.

Equation (\ref{spectrum1}) coincides with the spectrum in the
presence of normal scatterers \cite{been2,bagwellimp}. Note that
the gap in a symmetric contact Eq.\ (\ref{gap-sym}) vanishes for
certain phase difference $k_zd-\varphi_r=\pi n$. This is a result
of resonant tunneling through a system of two barriers with equal
reflection coefficients. The transmission probability $|T|^2$
through such system is unity at the resonance such that the gap
$\epsilon_g= 1-|T|^2$ disappears. A small asymmetry in the
scattering conditions removes the resonant tunneling effects so
that a gap will exist for any phase shift $k_zd-\varphi_r$. The
extreme asymmetric case is illustrated by Eq.\ (\ref{spectrum1}).
Similar effects of resonant tunneling and minigap oscillations as
functions of $k_zd$ can also take place for other mechanisms of
normal reflection such as interface barriers, mismatch in the
material parameters, or small normal scattering from the step-like
gap potential \cite{resonant}.

The predicted spectrum can be tuned by varying the magnetic field.
The minigap is not small and can reach values of the order of
$\Delta _0$ for $B_0\sim H_{cm}$. It can be monitored by measuring
the Josephson critical current that decreases with the minigap
\cite{bagwellimp}. Moreover, the minigap affects the dynamic
properties of the point contact. In particular, it is responsible
for suppression of the time-averaged quasiparticle current for
voltage biased contacts in the region $eV<\epsilon_g$ \cite{IV}.
Varying the magnetic field one can thus observe a transition from
ballistic to high-resistance behavior of the contact.


We thank I.A.\ Shereshevskii for stimulating discussions. This
work was supported in part by the US DOE Office of Science under
contract No. W-31-109-ENG-38, by Russian Foundation for Basic
Research, the Program ``Quantum Macrophysics'' of the Russian
Academy of Sciences, Russian State Fellowship for young doctors of
science (grant No. MD-141.2003.02), and Russian Science Support
Foundation. A.S.M. is grateful to the Low Temperature Laboratory
at the Helsinki University of Technology for hospitality.


\begin{thebibliography}{99}

\bibitem{galaiko}
V.\ P.\ Galaiko, Zh.\ Eksp.\ Teor.\ Fiz.\ {\bf 57}, 941 (1969)
[Sov.\ Phys.\ JETP\ {\bf 30}, 514 (1970)];
V.\ P.\ Galaiko and E.\ V.\ Bezuglyi, Zh.\ Eksp.\ Teor.\ Fiz.\
{\bf 60}, 1471 (1971) [Sov.\ Phys.\ JETP\ {\bf 33}, 796 (1971)];
G.\ A.\ Gogadze and I.\ O.\ Kulik, Zh.\ Eksp.\ Teor.\ Fiz.\ {\bf
60}, 1819 (1971) [Sov.\ Phys.\ JETP\ {\bf 33}, 984 (1971)].
\bibitem{tsoi}
S.\ I.\ Bozhko, V.\ S.\ Tsoi, and S.\ E.\ Yakovlev, Pis'ma\ Zh.\
Eksp.\ Teor.\ Fiz.\ {\bf 36}, 123 (1982) [JETP\ Lett.\ {\bf 36},
153 (1982)];
P.\ A.\ M.\ Benistant, A.\ P.\ van\ Gelder, H.\ van\ Kempen, and
P.\ Wyder, Phys.\ Rev.\ B\ {\bf 32}, 3351 (1985);
E.\ V.\ Bezuglyi and S.\ A.\ Korzh, Fiz.\ Nizk.\ Temp.\ {\bf 8},
420 (1982) [Sov. J. Low Temp. Phys. {\bf 8}, 208 (1982)].

\bibitem{kummel-v}
B.\ G\"otzelmann, S.\ Hofmann, and R.\ K\"ummel, Phys.\ Rev.\ B\
{\bf 52}, R3848 (1995).

\bibitem{Cook} P.\ M.\ A.\ Cook, V.\ C.\ Hui, and C.\ J.\ Lambert,
Europhys. Lett. {\bf 30}, 355 (1995).

\bibitem{interferometer}
H.\ Nakano and H.\ Takayanagi, Phys. Rev. B, {\bf 47}, 7986
(1993); A.\ F.\ Morpurgo, S.\ Holl, B.\ J.\ van\ Wees, and T.\ M.\
Klapwijk, Phys.\ Rev.\ Lett.\ {\bf 78}, 2636 (1997).

\bibitem{kulik-om1}
I.\ O.\ Kulik and A.\ N.\ Omel'yanchuk, Fiz.\ Nizk.\ Temp.\ {\bf
3}, 945 (1977) [Sov.\ J.\ Low\ Temp.\ Phys.\ {\bf 3}, 459 (1977)].

\bibitem{been1}
C.\ W.\ J.\ Beenakker and H.\ van\ Houten, Phys.\ Rev.\ Lett.\
{\bf 66}, 3056 (1991).

\bibitem{haber}
W.\ Habercorn, H.\ Knauer, and J.\ Richter, Phys.\ Status\ Solidi\
A\ {\bf 47}, K161 (1978).

\bibitem{zaitsev}
A.\ V.\ Zaitsev, Zh.\ Eksp.\ Teor.\ Fiz.\ {\bf 86}, 1742 (1984)
[Sov.\ Phys.\ JETP\ {\bf 59}, 1015 (1984)].

\bibitem{been2}
C.\ W.\ J.\ Beenakker, Phys.\ Rev.\ Lett.\ {\bf 67}, 3836 (1991).

\bibitem{bagwellimp}
P.\ F.\ Bagwell, Phys.\ Rev.\ B\ {\bf 46}, 12573 (1992).

\bibitem{BTK} G.E. Blonder, M. Tinkham, and T.M.Klapwijk, Phys.
Rev. B. {\bf 25}, 4515 (1982).

\bibitem{resonant}
A.\ L.\ Gudkov, M.\ Y.\ Kupriyanov, K.\ K.\ Likharev, Zh.\ Eksp.\
Teor.\ Fiz.\ {\bf 94}, 319 (1988) [Sov.\ Phys.\ JETP\ {\bf 68},
1478 (1988)];
H.\ van\ Houten, Appl.\ Phys.\ Lett.\ {\bf 58}, 1326 (1991);
U.\ Sch\"ussler and R.\ K\"ummel, Phys.\ Rev.\ B\ {\bf 47}, 2754
(1993);
M.\ Hurd and G.\ Wendin, Phys.\ Rev.\ B\ {\bf 49}, 15258 (1994).

\bibitem{IV} D. Averin and A. Bardas, Phys. Rev. Lett.
{\bf 75}, 1831 (1995); J.C. Cuevas, A. Mart{\'i}n-Rodero, and A. Levy
Yeyati, Phys. Rev. B {\bf 54}, 7366 (1996);
E. Scheer, P. Joyez, D. Esteve, C. Urbina, and
M.H. Devoret, Phys. Rev. Lett. {\bf 78}, 3535 (1997).








\end{thebibliography}
\end{document}